# Fluctuation-Electromagnetic Interaction between Rotating Nanoparticles. 2.Relativistic Theory

A. A. Kyasov and G. V. Dedkov

Nanoscale Physics Group, Kabardino-Balkarian State University, Nalchik, 360004, Russia

We study the fluctuation electromagnetic interaction in a system of two rotating electrically neutral nonmagnetic particles with allowance for relativistic retardation effect. The particles are assumed to have different temperatures being embedded into a heated vacuum background (photon gas). Closed expressions for the free energy, frictional torque and the rate of heat transfer are derived and compared to the known static results and those in the electrostatic limit without retardation.

## 1. Introduction

The forces between atoms, molecules and electrically neutral objects are often due to quantum and thermal fluctuations. These can be thought of as fluctuations of the molecular or atomic electric-dipole moments themselves, either permanent or induced, or as fluctuations in the electromagnetic fields surrounding the dipoles [1-3]. Radiative heat transfer between the bodies and nanoparticles at different temperatures has the same origin [4,5]. Quite recently, the fluctuation-electromagnetic effects for nanoparticles rotating in photon gas (vacuum background) [6], near a surface [7--9] and in a system of two spinning nanoparticles [10] have attracted great attention. In addition to the radiative heat transfer and frictional torque of rotating particles, we also calculated the interaction forces in the systems particle-surface and particle-particle [8--10]. Physically, the rotation of the particles leads to a change in the oscillation frequency of the electromagnetic field, so the force of attraction (van –der –Waals or Casimir –Polder) and the heat transfer differ from the corresponding values in the static case. Frictional torque is the net result of rotation, being analogous to the dissipative force in the case where a particle is moving at a constant velocity near the surface. This work is a natural extension of [10] with allowance for the retardation effects. We calculate the free energy, frictional torque and the rate of heat transfer of rotating particle in two geometric configurations (Figs. 1,2) assuming the second particle to be at rest. Essentially new is also the presence of a heated vacuum background (photon gas).

## 2. Free energy in a system of two rotating particles

First, we consider a geometrical and thermal configuration shown in Fig. 1, assuming the particles to be nonmagnetic. Both spherical particles 1 and 2 are assumed to be isotropic and are separated by the distance $R$. They are characterized by the frequency-dependent electric polarizabilities $\alpha_{1,2}(\omega)$ and different temperatures $T_1$, $T_2$, being embedded into the vacuum background with the temperature $T_3$. Without loss of generality, we can assume that the second particle is at rest with $\Sigma(X,Y,Z)$ denoting the corresponding reference frame. Another coordinate system $\Sigma'(x,y,z)$ rigidly rotates with the first particle with the angular velocity $\Omega$ around the $Z'(Z)$ axis (Fig. 1) that coincides with the vector $\mathbf{R}$. The free energy of the system depending on the distance $R$ is given by

$$U(R) = -\frac{1}{2}\langle \mathbf{d}_1(t)\mathbf{E}(\mathbf{r}_1,t)\rangle \tag{1}$$

where $\mathbf{d}_1(t)$ is the total fluctuation dipole moment of the first particle in the resting reference frame $\Sigma(x,y,z)$, that includes both spontaneous and induced (by the second particle and the vacuum background) contributions, $\mathbf{E}(\mathbf{r}_1,t)$ is the total fluctuation electromagnetic field at the location point $\mathbf{r}_1$ of the first particle relative to the system $\Sigma(x,y,z)$. It is worth noting that we must subtract from (1) the own energy of the fluctuation electromagnetic field that is not related to the interparticle interaction and which is obtained from (1) in the limit $R \to \infty$.

Within the general relativistic consideration of this problem, the sources of spontaneous fluctuations are the quantities $\mathbf{d}_i^{sp}, \mathbf{m}_i^{sp}, \mathbf{E}^{sp}, \mathbf{B}^{sp}$ ($i=1,2$), where the first two represent the fluctuating electric and magnetic moments of particles, and the last two represent the fluctuating spontaneous electric and magnetic fields of vacuum background. In what follows, since we aim to consider the case of nonmagnetic particles, the corresponding "magnetic" contributions to the interaction are omitted.

Using the frequency Fourier representation for $\mathbf{d}_1(t)$ and $\mathbf{E}(\mathbf{r}_1,t)$, Eq. (1) takes the form

$$U(R) = -\frac{1}{2}\int_{-\infty}^{+\infty}\frac{d\omega}{2\pi}\frac{d\omega'}{2\pi}\exp(-i(\omega+\omega')t)\langle d_{1,i}(\omega)E_i(\mathbf{r}_1,\omega')\rangle \tag{2}$$

In order to find the contributions to $d_{1,i}(\omega)$ produced by spontaneous dipole moments of the particles and the field $E_i(\mathbf{r}_1,\omega)$, and the contributions to $d_{1,i}(\omega)$ and $E_i(\mathbf{r}_1,\omega)$ produced by spontaneous fluctuations of vacuum, we need to obtain the relations between the induced dipole moment of the first (rotating) particle in $\Sigma(x,y,z)$ and the external field $E_i(\mathbf{r}_1,\omega)$ which is also taken in $\Sigma(x,y,z)$. The corresponding expressions are given by [9]

$$d^{in}_{1,x}(\omega) = \frac{1}{2}\left[\alpha_1(\omega_+)\left(E_x(\mathbf{r}_1,\omega) + iE_y(\mathbf{r}_1,\omega)\right) + \alpha_1(\omega_-)\left(E_x(\mathbf{r}_1,\omega) - iE_y(\mathbf{r}_1,\omega)\right)\right] \quad (3)$$

$$d^{in}_{1,y}(\omega) = \frac{1}{2}\left[-i\alpha_1(\omega_+)\left(E_x(\mathbf{r}_1,\omega) + iE_y(\mathbf{r}_1,\omega)\right) + i\alpha_1(\omega_-)\left(E_x(\mathbf{r}_1,\omega) - iE_y(\mathbf{r}_1,\omega)\right)\right] \quad (4)$$

$$d^{in}_{1,z}(\omega) = \alpha_1(\omega)E_z(\mathbf{r}_1,\omega),$$
$$\omega_\pm = \omega \pm \Omega \quad (5)$$

The components of the external field $E_i(\mathbf{r},\omega)$ in (3)—(5) are taken in $\Sigma(x,y,z)$ while $E_i(\mathbf{r},\omega)$ represents a sum of the vacuum field, spontaneous field of the second particle and induced field of the second particle created by the vacuum field. All these fields should be taken at $\mathbf{r}_1$. Using (3)—(5), the contributions of spontaneous dipole moments of particles in $d_{1,i}(\omega)$ are given by

$$d_{1,x}(\omega) = d^{sp}_{1,x}(\omega) + \left(-\frac{\omega^2}{\hbar c^2}\right)\cdot\frac{1}{2}\begin{Bmatrix}\alpha_1(\omega_+)\left(D_{xk}(\omega,\mathbf{R}) + iD_{yk}(\omega,\mathbf{R})\right)d^{sp}_{2,k}(\omega) + \\ \alpha_1(\omega_-)\left(D_{xk}(\omega,\mathbf{R}) - iD_{yk}(\omega,\mathbf{R})\right)d^{sp}_{2,k}(\omega)\end{Bmatrix} \quad (6)$$

$$d_{1,y}(\omega) = d^{sp}_{1,y}(\omega) + \left(-\frac{\omega^2}{\hbar c^2}\right)\cdot\frac{1}{2}\begin{Bmatrix}-i\alpha_1(\omega_+)\left(D_{xk}(\omega,\mathbf{R}) + iD_{yk}(\omega,\mathbf{R})\right)d^{sp}_{2,k}(\omega) + \\ +i\alpha_1(\omega_-)\left(D_{xk}(\omega,\mathbf{R}) - iD_{yk}(\omega,\mathbf{R})\right)d^{sp}_{2,k}(\omega)\end{Bmatrix} \quad (7)$$

$$d_{1,z}(\omega) = d^{sp}_{1,z}(\omega) + \left(-\frac{\omega^2}{\hbar c^2}\right)D_{zk}(\omega,\mathbf{R})d^{sp}_{2,k}(\omega) \quad (8)$$

where $D_{ik}(\omega,\mathbf{R})$ are the components of retarded Green's function of photons in free space [2] ($\mathbf{n} = \mathbf{R}/R$, $i,k = x,y,z$)

$$D_{ik}(\omega,\mathbf{R}) = \left(-\frac{\omega^2}{\hbar c^2}\right)\exp(i\omega R/c)\left[\left(\frac{\omega^2}{c^2 R} + \frac{i\omega}{cR} - \frac{1}{R^3}\right)(\delta_{ik} - n_i n_k) + 2\left(\frac{1}{R^3} - \frac{i\omega}{cR^2}\right)n_i n_k\right] \quad (9)$$

The first terms in (6)—(8) describe the own (spontaneous) dipole moments of the first particle, while the second terms describe the induced dipole moments created by the field of the second dipole. At the same time, the contributions from spontaneous dipole moments of particles in $E_i(\mathbf{r}_1,\omega)$ are given by

$$E_x(\mathbf{r}_1,\omega) = \left(-\frac{\omega^2}{\hbar c^2}\right)D_{xj}(\omega,\mathbf{R})d^{sp}{}_{2,j}(\omega) + \left(-\frac{\omega^2}{\hbar c^2}\right)^2 D_{xj}(\omega,\mathbf{R})D_{jl}(\omega,\mathbf{R})\alpha_2(\omega)d^{sp}{}_{1,l}(\omega) \quad (10)$$

$$E_y(\mathbf{r}_1,\omega) = \left(-\frac{\omega^2}{\hbar c^2}\right)D_{yj}(\omega,\mathbf{R})d^{sp}{}_{2,j}(\omega) + \left(-\frac{\omega^2}{\hbar c^2}\right)^2 D_{yj}(\omega,\mathbf{R})D_{jl}(\omega,\mathbf{R})\alpha_2(\omega)d^{sp}{}_{1,l}(\omega) \quad (11)$$

$$E_z(\mathbf{r}_1,\omega) = \left(-\frac{\omega^2}{\hbar c^2}\right)D_{zj}(\omega,\mathbf{R})d^{sp}{}_{2,j}(\omega) + \left(-\frac{\omega^2}{\hbar c^2}\right)^2 D_{zj}(\omega,\mathbf{R})D_{jl}(\omega,\mathbf{R})\alpha_2(\omega)d^{sp}{}_{1,l}(\omega) \quad (12)$$

The first terms in (10)—(12) describe the field of the second dipole at $\mathbf{r}_1$, while the second terms describe the fields of the second dipole induced by the first dipole.

The contributions of the fluctuating vacuum field in $d_{1,i}(\omega)$ and $E_i(\mathbf{r}_1,\omega)$ are given by

$$d_{1,x}(\omega) = \frac{1}{2}\left[\alpha_1(\omega_+)\left(E^{sp}{}_x(\mathbf{r}_1,\omega) + iE^{sp}{}_y(\mathbf{r}_1,\omega)\right) + \alpha_1(\omega_-)\left(E^{sp}{}_x(\mathbf{r}_1,\omega) - iE^{sp}{}_y(\mathbf{r}_1,\omega)\right)\right] + $$
$$+\frac{1}{2}\left(-\frac{\omega^2}{\hbar c^2}\right)\alpha_2(\omega)\begin{cases}\alpha_1(\omega_+)[D_{xk}(\omega,\mathbf{R}) + iD_{yk}(\omega,\mathbf{R})]E^{sp}{}_k(\mathbf{r}_2,\omega) + \\ + \alpha_1(\omega_-)[D_{xk}(\omega,\mathbf{R}) - iD_{yk}(\omega,\mathbf{R})]E^{sp}{}_k(\mathbf{r}_2,\omega)\end{cases} \quad (13)$$

$$d_{1,y}(\omega) = \frac{1}{2}\left[-i\alpha_1(\omega_+)\left(E^{sp}{}_x(\mathbf{r}_1,\omega) + iE^{sp}{}_y(\mathbf{r}_1,\omega)\right) + i\alpha_1(\omega_-)\left(E^{sp}{}_x(\mathbf{r}_1,\omega) - iE^{sp}{}_y(\mathbf{r}_1,\omega)\right)\right] + $$
$$+\frac{1}{2}\left(-\frac{\omega^2}{\hbar c^2}\right)\alpha_2(\omega)\begin{cases}-i\alpha_1(\omega_+)[D_{xk}(\omega,\mathbf{R}) + iD_{yk}(\omega,\mathbf{R})]E^{sp}{}_k(\mathbf{r}_2,\omega) + \\ + i\alpha_1(\omega_-)[D_{xk}(\omega,\mathbf{R}) - iD_{yk}(\omega,\mathbf{R})]E^{sp}{}_k(\mathbf{r}_2,\omega)\end{cases} \quad (14)$$

$$d_{1,z} = \alpha_1(\omega)E^{sp}{}_z(\mathbf{r}_1,\omega) + \left(-\frac{\omega^2}{\hbar c^2}\right)\alpha_2(\omega)\alpha_1(\omega)D_{zk}(\omega,\mathbf{R})E^{sp}{}_k(\mathbf{r}_2,\omega) \quad (15)$$

$$E_x(\mathbf{r}_1,\omega) = E^{sp}{}_x(\mathbf{r}_1,\omega) + \left(-\frac{\omega^2}{\hbar c^2}\right)D_{xk}(\omega,\mathbf{R})\alpha_2(\omega)E^{sp}{}_k(\mathbf{r}_2,\omega) \quad (16)$$

$$E_y(\mathbf{r}_1,\omega) = E^{sp}{}_y(\mathbf{r}_1,\omega) + \left(-\frac{\omega^2}{\hbar c^2}\right)D_{yk}(\omega,\mathbf{R})\alpha_2(\omega)E^{sp}{}_k(\mathbf{r}_2,\omega) \quad (17)$$

$$E_z(\mathbf{r}_1,\omega) = E^{sp}{}_z(\mathbf{r}_1,\omega) + \left(-\frac{\omega^2}{\hbar c^2}\right)D_{zk}(\omega,\mathbf{R})\alpha_2(\omega)E^{sp}{}_k(\mathbf{r}_2,\omega) \quad (18)$$

When inserting Eqs. (6)—(18) into (2), we must use the fluctuation-dissipation relations between components of the dipole moment and the electric field with allowance for the particle rotation relative to $\Sigma(x,y,z)$. These relations are derived in Appendix A and have the form (it is worth noting that the double-primed polarizabilities $\alpha$ denote their imaginary components)

$$\langle d^{sp}{}_{1,x}(\omega)d^{sp}{}_{1,x}(\omega')\rangle = \langle d^{sp}{}_{1,y}(\omega)d^{sp}{}_{1,y}(\omega')\rangle = \frac{1}{2}2\pi\hbar\delta(\omega+\omega')\cdot$$
$$\cdot\left[\alpha_1''(\omega_+)\coth\frac{\hbar\omega_+}{2k_BT_1}+\alpha_1''(\omega_-)\coth\frac{\hbar\omega_-}{2k_BT_2}\right] \quad (19)$$

$$\langle d^{sp}{}_{1,x}(\omega)d^{sp}{}_{1,y}(\omega')\rangle = -\langle d^{sp}{}_{1,y}(\omega)d^{sp}{}_{1,x}(\omega')\rangle = \frac{i}{2}2\pi\hbar\delta(\omega+\omega')\cdot$$
$$\cdot\left[\alpha_1''(\omega_+)\coth\frac{\hbar\omega_+}{2k_BT_1}-\alpha_1''(\omega_-)\coth\frac{\hbar\omega_-}{2k_BT_1}\right] \quad (20)$$

$$\langle d^{sp}{}_{1,x}(\omega)d^{sp}{}_{1,z}(\omega')\rangle = \langle d^{sp}{}_{1,y}(\omega)d^{sp}{}_{1,z}(\omega')\rangle = 0 \quad (21)$$

$$\langle d^{sp}{}_{2,i}(\omega)d^{sp}{}_{2,k}(\omega')\rangle = 2\pi\hbar\delta_{ik}\delta(\omega+\omega')\alpha_2''(\omega)\coth\frac{\hbar\omega}{2k_BT_2} \quad (22)$$

$$\langle E^{sp}{}_i(\mathbf{r}_1,\omega)E^{sp}{}_k(\mathbf{r}_2,\omega')\rangle = 2\pi\hbar\delta(\omega+\omega')\left(-\frac{\omega^2}{\hbar c^2}\right)\mathrm{Im}\,D_{ik}(\omega,\mathbf{R})\coth\frac{\hbar\omega}{2k_BT_3} \quad (23)$$

Using (6)—(23), after straightforward algebra Eq. (2) takes the form ($\mu = x, y$).

$$U(R) = -\frac{\hbar}{2\pi}\int_0^\infty d\omega\left(\frac{\omega^2}{\hbar c^2}\right)^2\cdot$$
$$\left\{\begin{array}{l}\mathrm{Re}[D_{iz}(\omega,\mathbf{R})D_{iz}(\omega,\mathbf{R})\alpha_2(\omega)]\alpha_1''(\omega)\coth\dfrac{\hbar\omega}{2k_BT_1}+\dfrac{1}{2}\mathrm{Re}[D_{i\mu}(\omega,\mathbf{R})D_{i\mu}(\omega,\mathbf{R})\alpha_2(\omega)]\cdot \\[6pt]
\cdot\left[\alpha_1''(\omega_+)\coth\dfrac{\hbar\omega_+}{2k_BT_1}+\alpha_1''(\omega_-)\coth\dfrac{\hbar\omega_-}{2k_BT_1}\right]+ \\[6pt]
+\mathrm{Re}[D^*{}_{iz}(\omega,\mathbf{R})D_{iz}(\omega,\mathbf{R})\alpha_1(\omega)]\alpha_2''(\omega)\coth\dfrac{\hbar\omega}{2k_BT_2}+ \\[6pt]
\dfrac{1}{2}\mathrm{Re}[D^*{}_{i\mu}(\omega,\mathbf{R})D_{i\mu}(\omega,\mathbf{R})(\alpha_1(\omega_+)+\alpha_1(\omega_-))]\cdot\alpha_2''(\omega)\coth\dfrac{\hbar\omega}{2k_BT_2}+ \\[6pt]
+\mathrm{Re}[\alpha_1^*(\omega)\alpha_2(\omega)D_{iz}(\omega,\mathbf{R})]D_{iz}''(\omega,\mathbf{R})\coth\dfrac{\hbar\omega}{2k_BT_3}+ \\[6pt]
+\dfrac{1}{2}\mathrm{Re}[(\alpha_1^*(\omega_+)+\alpha_1^*(\omega_-))\alpha_2(\omega)D_{i\mu}(\omega,\mathbf{R})]\cdot D_{i\mu}''(\omega,\mathbf{R})\coth\dfrac{\hbar\omega}{2k_BT_3}+ \\[6pt]
+\mathrm{Re}[\alpha_1(\omega)\alpha_2(\omega)D_{iz}(\omega,\mathbf{R})]D_{iz}''(\omega,\mathbf{R})\coth\dfrac{\hbar\omega}{2k_BT_3}+ \\[6pt]
+\dfrac{1}{2}\mathrm{Re}[(\alpha_1(\omega_+)+\alpha_1(\omega_-))\alpha_2(\omega)D_{i\mu}(\omega,\mathbf{R})]\cdot D_{i\mu}''(\omega,\mathbf{R})\coth\dfrac{\hbar\omega}{2k_BT_3}\end{array}\right\} \quad (24)$$

Note that complex conjugated values are marked by the sign *, real and imaginary parts are denoted by one and two primes.

Let us consider some limiting cases of Eq. (24). In the static case $\Omega = 0$, after some rearrangements with allowance for parity (oddness) of real and imaginary parts of the functions $\alpha_{1,2}(\omega), D_{ik}(\omega, \mathbf{R})$ we obtain

$$U(R) = -\frac{\hbar}{2\pi} \int_0^\infty d\omega \left(\frac{\omega^2}{\hbar c^2}\right)^2 \cdot$$

$$\cdot \left\{ \begin{array}{l} \operatorname{Re}[D_{ik}(\omega, \mathbf{R}) D_{ik}(\omega, \mathbf{R})] \operatorname{Im}[\alpha_1(\omega)\alpha_2(\omega)] \coth \dfrac{\hbar\omega}{2k_B T_1} + \\[6pt] + \operatorname{Im}[D_{ik}(\omega, \mathbf{R}) D_{ik}(\omega, \mathbf{R})] \operatorname{Re}[\alpha_1(\omega)\alpha_2(\omega)] \coth \dfrac{\hbar\omega}{2k_B T_3} + \\[6pt] 2 D'_{ik}(\omega, \mathbf{R}) D''_{ik}(\omega, \mathbf{R}) \alpha''_1(\omega)\alpha''_2(\omega) \left[\coth \dfrac{\hbar\omega}{2k_B T_3} - \coth \dfrac{\hbar\omega}{2k_B T_1}\right] + \\[6pt] + D'_{ik}(\omega, \mathbf{R}) D'_{ik}(\omega, \mathbf{R}) \alpha'_1(\omega)\alpha''_2(\omega) \left[\coth \dfrac{\hbar\omega}{2k_B T_2} - \coth \dfrac{\hbar\omega}{2k_B T_1}\right] + \\[6pt] + D''_{ik}(\omega, \mathbf{R}) D''_{ik}(\omega, \mathbf{R}) \alpha'_1(\omega)\alpha''_2(\omega) \left[\coth \dfrac{\hbar\omega}{2k_B T_1} + \coth \dfrac{\hbar\omega}{2k_B T_2} - 2\coth \dfrac{\hbar\omega}{2k_B T_3}\right] \end{array} \right\}$$

(25)

In the case of total thermal equilibrium $T_1 = T_2 = T_3$ Eq. (25) takes the form

$$U(R) = -\frac{\hbar}{2\pi} \operatorname{Im} \left[\int_0^\infty d\omega \left(\frac{\omega^2}{\hbar c^2}\right)^2 \coth \frac{\hbar\omega}{2k_B T} \alpha_1(\omega)\alpha_2(\omega) D_{ik}(\omega, \mathbf{R}) D_{ik}(\omega, \mathbf{R})\right] \qquad (26)$$

From (26) at $T = 0$, using the standard transformation of the integral in the complex frequency plane, we obtain the classical result by Casimir and Polder [1] (see also [11])

$$U(R) = -\frac{\hbar}{\pi R^6} \int_0^\infty d\omega \, \alpha_1(i\omega)\alpha_2(i\omega) \exp(-2\omega R/c) \left(\frac{\omega R}{c}\right)^4 \cdot$$

$$\cdot \left[1 + 2\left(\frac{\omega R}{c}\right)^{-1} + 5\left(\frac{\omega R}{c}\right)^{-2} + 6\left(\frac{\omega R}{c}\right)^{-3} + 3\left(\frac{\omega R}{c}\right)^{-4}\right]$$

(27)

An important nontrivial consequence of Eq. (25) is the lack of the symmetry with respect to the transformations $\alpha_1 \leftrightarrow \alpha_2, T_1 \leftrightarrow T_2$ if $T_1 \neq T_2 \neq T_3$. This means that the force acting on the first particle and the force acting on the second one are different, while the uncompensated force should be applied to the vacuum background. This is true in the particular case where the retardation effect is present. In other words, two electrically neutral particles having different temperature violate the symmetry of vacuum regarding the Casimir force. This is in line with the result [12] obtained in the case of two thick interacting slabs out of thermal equilibrium.

## 3. Heating(cooling) rate of rotating particle

The starting equation for the heating rate $\dot{Q}$ in $\Sigma_2(x,y,z)$ is given by

$$\dot{Q} = \langle \dot{\mathbf{d}}_1(t)\mathbf{E}(\mathbf{r}_1,t)\rangle = \int_{-\infty}^{+\infty}\frac{d\omega\, d\omega'}{2\pi\, 2\pi}(-i\omega)\exp(-i(\omega+\omega')t)\langle d_{1,i}(\omega)E_i(\mathbf{r}_1,\omega')\rangle \tag{28}$$

In contrast to Eq. (2), the total electric field in (28) taken at the point $\mathbf{r}_1$ should include the vacuum field $\mathbf{E}^{in}$ that is induced by spontaneous dipole moment of the first particle. This field did not result in free energy (2).

Making the calculations similar to those we made in deriving Eq. (25) we finally obtain

$$\dot{Q} = -\frac{2\hbar}{3\pi c^3}\int_0^\infty d\omega\,\omega^4 \left\{\begin{array}{l}\alpha_1''(\omega)\left[\coth\dfrac{\hbar\omega}{2k_B T_1} - \coth\dfrac{\hbar\omega}{2k_B T_3}\right] + \\ \alpha_1''(\omega_+)\left[\coth\dfrac{\hbar\omega_+}{2k_B T_1} - \coth\dfrac{\hbar\omega}{2k_B T_3}\right] + \\ \alpha_1''(\omega_-)\left[\coth\dfrac{\hbar\omega_-}{2k_B T_1} - \coth\dfrac{\hbar\omega}{2k_B T_3}\right]\end{array}\right\} - \frac{\hbar}{\pi}\int_0^\infty d\omega\,\omega\left(\frac{\omega^2}{\hbar c^2}\right)\cdot$$

$$\cdot\left\{\begin{array}{l}\mathrm{Im}[D_{iz}(\omega,\mathbf{R})D_{iz}(\omega,\mathbf{R})\alpha_2(\omega)]\alpha_1''(\omega)\coth\dfrac{\hbar\omega}{2k_B T_1} + \dfrac{1}{2}\mathrm{Im}[D_{i\mu}(\omega,\mathbf{R})D_{i\mu}(\omega,\mathbf{R})\alpha_2(\omega)]\cdot \\ \cdot\left[\alpha_1''(\omega_+)\coth\dfrac{\hbar\omega_+}{2k_B T_1} + \alpha_1''(\omega_-)\coth\dfrac{\hbar\omega_-}{2k_B T_1}\right] - \\ -\mathrm{Im}[D^*_{iz}(\omega,\mathbf{R})D_{iz}(\omega,\mathbf{R})\alpha_1(\omega)]\alpha_2''(\omega)\coth\dfrac{\hbar\omega}{2k_B T_2} - \\ -\dfrac{1}{2}\mathrm{Im}[D^*_{i\mu}(\omega,\mathbf{R})D_{i\mu}(\omega,\mathbf{R})(\alpha_1(\omega_+)+\alpha_1(\omega_-))]\cdot\alpha_2''(\omega)\coth\dfrac{\hbar\omega}{2k_B T_2} + \\ +\mathrm{Im}[\alpha_1^*(\omega)\alpha_2(\omega)D_{iz}(\omega,\mathbf{R})]D_{iz}''(\omega,\mathbf{R})\coth\dfrac{\hbar\omega}{2k_B T_3} + \\ +\dfrac{1}{2}\mathrm{Im}[(\alpha_1^*(\omega_+)+\alpha_1^*(\omega_-))\alpha_2(\omega)D_{i\mu}(\omega,\mathbf{R})]\cdot D_{i\mu}''(\omega,\mathbf{R})\coth\dfrac{\hbar\omega}{2k_B T_3} - \\ -\mathrm{Im}[\alpha_1(\omega)\alpha_2(\omega)D_{iz}(\omega,\mathbf{R})]D_{iz}''(\omega,\mathbf{R})\coth\dfrac{\hbar\omega}{2k_B T_3} - \\ -\dfrac{1}{2}\mathrm{Im}[(\alpha_1(\omega_+)+\alpha_1(\omega_-))\alpha_2(\omega)D_{i\mu}(\omega,\mathbf{R})]\cdot D_{i\mu}''(\omega,\mathbf{R})\coth\dfrac{\hbar\omega}{2k_B T_3}\end{array}\right\} \tag{29}$$

where we put again $\mu = x, z$. Equation (29) consists of two different terms. The first one does not depend on the distance $R$ and describes the heat exchange between rotating particle and vacuum irrespectively of the presence of the second particle. This contribution coincides with the result [6] (see also Appendix B). The second term describes the heat exchange between the two particles with allowance for additional correlation between the fluctuating dipole moments caused by vacuum field. In the static limit $\Omega = 0$, after simple transformations Eq. (29) takes the form

$$\dot{Q} = -\frac{2\hbar}{\pi c^3}\int_0^\infty d\omega\, \omega^4 \alpha_1''(\omega)\left[\coth\frac{\hbar\omega}{2k_B T_1} - \coth\frac{\hbar\omega}{2k_B T_3}\right] - \frac{\hbar}{\pi}\int_0^\infty d\omega\, \omega\left(\frac{\omega^2}{\hbar c^2}\right)^2 \cdot$$

$$\cdot \left\{ \begin{array}{l} \alpha_1''(\omega)\alpha_2''(\omega)|D_{ik}(\omega,\mathbf{R})|^2 \left[\coth\dfrac{\hbar\omega}{2k_B T_1} - \coth\dfrac{\hbar\omega}{2k_B T_2}\right] + \\ + 2\alpha_1''(\omega)D_{ik}''(\omega,\mathbf{R})\operatorname{Re}[\alpha_2(\omega)D_{ik}(\omega,\mathbf{R})]\cdot\left[\coth\dfrac{\hbar\omega}{2k_B T_1} - \coth\dfrac{\hbar\omega}{2k_B T_3}\right] \end{array}\right\} \quad (30)$$

Equation (30) exactly coincides with the result [6]. At $T_1 = T_3$ the first integral term of (30) and the vacuum-correlation part (the second term in the figure brackets) disappear. Then with allowance for (9) Eq. (30) takes the form [13]

$$\dot{Q} = -\frac{2\hbar}{\pi R^6}\int_0^\infty d\omega\, \omega\, \alpha_1''(\omega)\alpha_2''(\omega)\left(3 + \left(\frac{\omega R}{c}\right)^2 + \left(\frac{\omega R}{c}\right)^4\right)\left[\coth\frac{\hbar\omega}{2k_B T_1} - \coth\frac{\hbar\omega}{2k_B T_2}\right] \quad (31)$$

In the general case $T_1 \neq T_2 \neq T_3$ (in contrast to [13]), the correct expression for the vacuum-mediated thermal radiation of the first particle also includes the contribution being proportional to $\operatorname{Re}[\alpha_2(\omega)D_{ik}(\omega,\mathbf{R})]$.

## 4. Frictional torque

The torque produced by an electric field $\mathbf{E}$ on a dipole $\mathbf{d}$ is given by $\mathbf{d}\times\mathbf{E}$. In our case, the only one nonzero component of the torque has the form

$$M_z = \langle \mathbf{d}_1(t)\times\mathbf{E}(\mathbf{r}_1,t)\rangle_z = \int_{-\infty}^{+\infty}\int_{-\infty}^{+\infty}\frac{d\omega\, d\omega'}{2\pi\, 2\pi}\exp(-i(\omega+\omega')t)\cdot\langle d_{1,x}(\omega)E_y(\mathbf{r}_1,\omega') - d_{1,y}(\omega)E_x(\mathbf{r}_1,\omega')\rangle \quad (32)$$

In contrast to Eq. (2), we must also take in this case the field $\mathbf{E}(\mathbf{r}_1,\omega)$ with the additional contribution $\mathbf{E}^{in}(\mathbf{r}_1,\omega)$ --- the vacuum field that is induced by the spontaneous dipole moment of the first particle (as in deriving the expression for $\dot{Q}$). Other contributions in $\mathbf{E}(\mathbf{r}_1,\omega)$ are given by (10), (11), (16) and (17). The components $d_{1,x(y)}(\omega)$ are determined from (6), (7) and (13), (14). Calculating the corresponding correlators with allowance for (20), (22), (23) yields

$$M_z = -\frac{2\hbar}{3\pi c^3}\int_0^\infty d\omega \omega^3 \left\{ \begin{array}{l} \alpha_1''(\omega_-)\left[\coth\frac{\hbar\omega_-}{2k_BT_1} - \coth\frac{\hbar\omega_-}{2k_BT_3}\right] - \\ \alpha_1''(\omega_+)\left[\coth\frac{\hbar\omega_+}{2k_BT_1} - \coth\frac{\hbar\omega_+}{2k_BT_3}\right] + \end{array} \right\} - \frac{\hbar}{2\pi}\int_0^\infty d\omega \left(\frac{\omega^2}{\hbar c^2}\right)^2 \cdot$$

$$\left\{ \begin{array}{l} \mathrm{Im}\left[(D^2{}_{xx}(\omega,\mathbf{R}) + D^2{}_{yy}(\omega,\mathbf{R}))\alpha_2(\omega)\right]\cdot\left(\alpha_1''(\omega_-)\coth\frac{\hbar\omega_-}{2k_BT_1} - \alpha_1''(\omega_+)\coth\frac{\hbar\omega_+}{2k_BT_1}\right) + \\ + \mathrm{Im}\left[(|D_{xx}(\omega,\mathbf{R})|^2 + |D_{yy}(\omega,\mathbf{R})|^2)\alpha_2(\omega)\right]\cdot(\alpha_1''(\omega_+) - \alpha_1''(\omega_-))\coth\frac{\hbar\omega}{2k_BT_2} + \\ + \left[2\mathrm{Re}(\alpha_2(\omega)D_{xx}(\omega,\mathbf{R}))\mathrm{Im}\,D_{xx}(\omega,\mathbf{R}) + 2\mathrm{Re}(\alpha_2(\omega)D_{yy}(\omega,\mathbf{R}))\mathrm{Im}\,D_{yy}(\omega,\mathbf{R})\right]\cdot \\ \cdot(\alpha_1''(\omega_+) - \alpha_1''(\omega_-))\coth\frac{\hbar\omega}{2k_BT_3} \end{array} \right\} \quad (33)$$

Similar to Eq. (30) for the rate of heating $Q$, the friction torque consists of three different parts. The first of them is given by the first integral and it is does not depend on the distance $R$. This part of torque corresponds to the frictional moment produced by the vacuum background irrespectively of the second particle [6] and in the limit $c \to \infty$ it equals zero. The second part (two first terms in figure brackets of the second integral has a nonzero nonrelativistic limit produced by the spontaneous dipole moments of particles, and the third part (the last term in figure brackets of the second integral) is due to the additional correlation of the dipole moments mediated by vacuum field. In the limit $c \to \infty$ this part also equals zero.

### 5. The case with the rotation axis parallel to the surface

In this case, the geometrical configuration and coordinate systems used are shown in Fig. 2. Performing the calculations quite analogous to the case shown in Fig.1, we obtain the expressions that differ from Eqs. (24), (29) and (33) by cyclic permutation $x \to y \to z \to x$. The resulting formulas have the form

$$U(R) = -\frac{\hbar}{2\pi} \int_0^\infty d\omega \left(\frac{\omega^2}{\hbar c^2}\right)^2 \cdot$$

$$\begin{cases}
\text{Re}[D_{ix}(\omega,\mathbf{R})D_{ix}(\omega,\mathbf{R})\alpha_2(\omega)]\alpha_1''(\omega)\coth\frac{\hbar\omega}{2k_BT_1} + \frac{1}{2}\text{Re}[D_{i\mu}(\omega,\mathbf{R})D_{i\mu}(\omega,\mathbf{R})\alpha_2(\omega)] \cdot \\
\cdot \left[\alpha_1''(\omega_+)\coth\frac{\hbar\omega_+}{2k_BT_1} + \alpha_1''(\omega_-)\coth\frac{\hbar\omega_-}{2k_BT_1}\right] + \\
+ \text{Re}[D^*_{ix}(\omega,\mathbf{R})D_{ix}(\omega,\mathbf{R})\alpha_1(\omega)]\alpha_2''(\omega)\coth\frac{\hbar\omega}{2k_BT_2} + \\
\frac{1}{2}\text{Re}[D^*_{i\mu}(\omega,\mathbf{R})D_{i\mu}(\omega,\mathbf{R})(\alpha_1(\omega_+)+\alpha_1(\omega_-))] \cdot \alpha_2''(\omega)\coth\frac{\hbar\omega}{2k_BT_2} + \\
+ \text{Re}[\alpha_1^*(\omega)\alpha_2(\omega)D_{ix}(\omega,\mathbf{R})]D_{ix}''(\omega,\mathbf{R})\coth\frac{\hbar\omega}{2k_BT_3} + \\
+ \frac{1}{2}\text{Re}[(\alpha_1^*(\omega_+)+\alpha_1^*(\omega_-))\alpha_2(\omega)D_{i\mu}(\omega,\mathbf{R})] \cdot D_{i\mu}''(\omega,\mathbf{R})\coth\frac{\hbar\omega}{2k_BT_3} + \\
+ \text{Re}[\alpha_1(\omega)\alpha_2(\omega)D_{ix}(\omega,\mathbf{R})]D_{ix}''(\omega,\mathbf{R})\coth\frac{\hbar\omega}{2k_BT_3} + \\
+ \frac{1}{2}\text{Re}[(\alpha_1(\omega_+)+\alpha_1(\omega_-))\alpha_2(\omega)D_{i\mu}(\omega,\mathbf{R})] \cdot D_{i\mu}''(\omega,\mathbf{R})\coth\frac{\hbar\omega}{2k_BT_3}
\end{cases} \quad (34)$$

$$\dot{Q} = -\frac{2\hbar}{3\pi c^3} \int_0^\infty d\omega \omega^4 \begin{Bmatrix} \alpha_1''(\omega)\left[\coth\frac{\hbar\omega}{2k_BT_1} - \coth\frac{\hbar\omega}{2k_BT_3}\right] + \\ \alpha_1''(\omega_+)\left[\coth\frac{\hbar\omega_+}{2k_BT_1} - \coth\frac{\hbar\omega}{2k_BT_3}\right] + \\ \alpha_1''(\omega_-)\left[\coth\frac{\hbar\omega_-}{2k_BT_1} - \coth\frac{\hbar\omega}{2k_BT_3}\right] \end{Bmatrix} - \frac{\hbar}{\pi}\int_0^\infty d\omega\omega\left(\frac{\omega^2}{\hbar c^2}\right)^2 \cdot$$

$$\cdot \begin{Bmatrix} \mathrm{Im}[D_{ix}(\omega,\mathbf{R})D_{ix}(\omega,\mathbf{R})\alpha_2(\omega)]\alpha_1''(\omega)\coth\frac{\hbar\omega}{2k_BT_1} + \frac{1}{2}\mathrm{Im}[D_{i\mu}(\omega,\mathbf{R})D_{i\mu}(\omega,\mathbf{R})\alpha_2(\omega)] \cdot \\ \cdot\left[\alpha_1''(\omega_+)\coth\frac{\hbar\omega_+}{2k_BT_1} + \alpha_1''(\omega_-)\coth\frac{\hbar\omega_-}{2k_BT_1}\right] - \\ - \mathrm{Im}[D^*_{ix}(\omega,\mathbf{R})D_{ix}(\omega,\mathbf{R})\alpha_1(\omega)]\alpha_2''(\omega)\coth\frac{\hbar\omega}{2k_BT_2} - \\ - \frac{1}{2}\mathrm{Im}[D^*_{i\mu}(\omega,\mathbf{R})D_{i\mu}(\omega,\mathbf{R})(\alpha_1(\omega_+) + \alpha_1(\omega_-))] \cdot \alpha_2''(\omega)\coth\frac{\hbar\omega}{2k_BT_2} + \\ + \mathrm{Im}[\alpha_1^*(\omega)\alpha_2(\omega)D_{ix}(\omega,\mathbf{R})]D_{ix}''(\omega,\mathbf{R})\coth\frac{\hbar\omega}{2k_BT_3} + \\ + \frac{1}{2}\mathrm{Im}[(\alpha_1^*(\omega_+) + \alpha_1^*(\omega_-))\alpha_2(\omega)D_{i\mu}(\omega,\mathbf{R})] \cdot D_{i\mu}''(\omega,\mathbf{R})\coth\frac{\hbar\omega}{2k_BT_3} - \\ - \mathrm{Im}[\alpha_1(\omega)\alpha_2(\omega)D_{ix}(\omega,\mathbf{R})]D_{ix}''(\omega,\mathbf{R})\coth\frac{\hbar\omega}{2k_BT_3} - \\ - \frac{1}{2}\mathrm{Im}[(\alpha_1(\omega_+) + \alpha_1(\omega_-))\alpha_2(\omega)D_{i\mu}(\omega,\mathbf{R})] \cdot D_{i\mu}''(\omega,\mathbf{R})\coth\frac{\hbar\omega}{2k_BT_3} \end{Bmatrix} \quad (35)$$

$$M_x = -\frac{2\hbar}{3\pi c^3}\int_0^\infty d\omega\omega^3 \begin{Bmatrix} \alpha_1''(\omega_-)\left[\coth\frac{\hbar\omega_-}{2k_BT_1} - \coth\frac{\hbar\omega}{2k_BT_3}\right] - \\ \alpha_1''(\omega_+)\left[\coth\frac{\hbar\omega_+}{2k_BT_1} - \coth\frac{\hbar\omega}{2k_BT_3}\right] + \end{Bmatrix} - \frac{\hbar}{2\pi}\int_0^\infty d\omega\left(\frac{\omega^2}{\hbar c^2}\right)^2 \cdot$$

$$\cdot \begin{Bmatrix} \mathrm{Im}[(D^2_{yy}(\omega,\mathbf{R}) + D^2_{zz}(\omega,\mathbf{R}))\alpha_2(\omega)] \cdot \left(\alpha_1''(\omega_-)\coth\frac{\hbar\omega_-}{2k_BT_1} - \alpha_1''(\omega_+)\coth\frac{\hbar\omega_+}{2k_BT_1}\right) + \\ + \mathrm{Im}[(|D_{yy}(\omega,\mathbf{R})|^2 + |D_{zz}(\omega,\mathbf{R})|^2)\alpha_2(\omega)] \cdot (\alpha_1''(\omega_+) - \alpha_1''(\omega_-))\coth\frac{\hbar\omega}{2k_BT_2} + \\ + [2\mathrm{Re}(\alpha_2(\omega)D_{yy}(\omega,\mathbf{R}))\mathrm{Im}\,D_{yy}(\omega,\mathbf{R}) + 2\mathrm{Re}(\alpha_2(\omega)D_{zz}(\omega,\mathbf{R}))\mathrm{Im}\,D_{zz}(\omega,\mathbf{R})] \cdot \\ \cdot (\alpha_1''(\omega_+) - \alpha_1''(\omega_-))\coth\frac{\hbar\omega}{2k_BT_3} \end{Bmatrix} \quad (36)$$

Moreover, one should bear in mind that in contrast to (24), (29), in Eqs. (34), (35) $\mu = y, z$

## 6. Nonrelativistic case

In the nonrelativistic case $c \to \infty$, in both configurations shown in Figs. 1,2 the expressions for $U(z_0), \dot{Q}, M_{z,x}$ were obtained in our paper [10]. Formulas (24), (29), (30) and (34)—(36) at $c \to \infty$ are in complete agreement with these results. The corresponding expressions have the form

a) $\mathbf{\Omega} = (0,0,\Omega)$

$$U = -\frac{\hbar}{2\pi R^6} \int_0^\infty d\omega \left[ 4\left(\alpha_1''(\omega)\alpha_2'(\omega)\coth\frac{\hbar\omega}{2k_B T_1} + \alpha_1'(\omega)\alpha_2''(\omega)\coth\frac{\hbar\omega}{2k_B T_2}\right) + \right.$$
$$+ \alpha_1''(\omega_+)\alpha_2'(\omega)\coth\frac{\hbar\omega_+}{2k_B T_1} + \alpha_1'(\omega_+)\alpha_2''(\omega)\coth\frac{\hbar\omega}{2k_B T_2} +$$
$$\left. + \alpha_1''(\omega_-)\alpha_2'(\omega)\coth\frac{\hbar\omega_-}{2k_B T_1} + \alpha_1'(\omega_-)\alpha_2''(\omega)\coth\frac{\hbar\omega}{2k_B T_2} \right] \quad (37)$$

$$\dot{Q} = \frac{\hbar}{\pi R^6} \int_0^\infty d\omega\, \omega\, \alpha_2''(\omega) \left[ 4\alpha_1''(\omega)\left(\coth\frac{\hbar\omega}{2k_B T_2} - \coth\frac{\hbar\omega}{2k_B T_1}\right) + \right.$$
$$\left. + \alpha_1''(\omega_+)\left(\coth\frac{\hbar\omega}{2k_B T_2} - \coth\frac{\hbar\omega_+}{2k_B T_1}\right) + \alpha_1''(\omega_-)\left(\coth\frac{\hbar\omega}{2k_B T_2} - \coth\frac{\hbar\omega_-}{2k_B T_1}\right) \right] \quad (38)$$

$$M_z = -\frac{\hbar}{\pi R^6}\int_0^\infty d\omega\, \alpha_2''(\omega) \left[ \begin{array}{l} \alpha_1''(\omega_-)\left(\coth\frac{\hbar\omega_-}{2k_B T_1} - \coth\frac{\hbar\omega}{2k_B T_2}\right) - \\ -\alpha_1''(\omega_+)\left(\coth\frac{\hbar\omega_+}{2k_B T_1} - \coth\frac{\hbar\omega}{2k_B T_2}\right) \end{array} \right] \quad (39)$$

b) $\mathbf{\Omega} = (\Omega,0,0)$

$$U = -\frac{\hbar}{4\pi R^6}\int_0^\infty d\omega \left[ 2\left(\alpha_1''(\omega)\alpha_2'(\omega)\coth\frac{\hbar\omega}{2k_B T_1} + \alpha_1'(\omega)\alpha_2''(\omega)\coth\frac{\hbar\omega}{2k_B T_2}\right) + \right.$$
$$+ 5\left(\alpha_1''(\omega_+)\alpha_2'(\omega)\coth\frac{\hbar\omega_+}{2k_B T_1} + \alpha_1'(\omega_+)\alpha_2''(\omega)\coth\frac{\hbar\omega}{2k_B T_2}\right) + \quad (40)$$
$$\left. + 5\left(\alpha_1''(\omega_-)\alpha_2'(\omega)\coth\frac{\hbar\omega_-}{2k_B T_1} + \alpha_1'(\omega_-)\alpha_2''(\omega)\coth\frac{\hbar\omega}{2k_B T_2}\right) \right]$$

$$\dot{Q} = \frac{\hbar}{2\pi R^6}\int_0^\infty d\omega\, \omega\, \alpha_2''(\omega) \left[ 2\alpha_1''(\omega)\left(\coth\frac{\hbar\omega}{2k_B T_2} - \coth\frac{\hbar\omega}{2k_B T_1}\right) + \right.$$
$$\left. + 5\alpha_1''(\omega_+)\left(\coth\frac{\hbar\omega}{2k_B T_2} - \coth\frac{\hbar\omega_+}{2k_B T_1}\right) + 5\alpha_1''(\omega_-)\left(\coth\frac{\hbar\omega}{2k_B T_2} - \coth\frac{\hbar\omega_-}{2k_B T_1}\right) \right] \quad (41)$$

$$M_x = -\frac{5\hbar}{2\pi R^6} \int_0^\infty d\omega\, \alpha_2''(\omega) \begin{bmatrix} \alpha_1''(\omega_-)\left(\coth\frac{\hbar\omega_-}{2k_B T_1} - \coth\frac{\hbar\omega}{2k_B T_2}\right) - \\ -\alpha_1''(\omega_+)\left(\coth\frac{\hbar\omega_+}{2k_B T_1} - \coth\frac{\hbar\omega}{2k_B T_2}\right) \end{bmatrix} \qquad (42)$$

## 7. Summary and conclusions

Using the general background of the fluctuation electromagnetic theory, we have obtained closed relativistic expressions for the free energy, frictional torque and the rate of heat transfer in a system of two rotating particles, characterizing by different temperatures and electric polarizabilities. Thermal state of vacuum (photon gas) is arbitrary.

The results previously obtained in the static limit or in the case without allowance for the retardation follow from the general formulas. It is worth noting that all formulas obtained in this paper can be also applied to the system of two rotating magnetically polarizable particles (without electrical polarization) provided that $\alpha \to \chi$ (where $\chi$ is the magnetic polarizability). In the case where the particles have both the electric and magnetic polarizabilities, one should bear in mind also the presence of the cross-terms $\alpha_1 \chi_2$ and vice versa.

An important conclusion is that even in the case of nonrotating particles out of thermal equilibrium between one another and with the vacuum background, the forces acting on the particles do not obey the Newton's third law.

# APPENDIX A

Fluctuation-dissipation relations for the rotating fluctuating dipole in the resting reference frame $\Sigma(X,Y,Z)$

a) $\mathbf{\Omega}=(0,0,\Omega)$ (Fig. 1)

Using the transformation formulas for the Fourier-components of the dipole moment of a rotating particle when passing from a resting to a rotating coordinate system [9]

$$d_x^{sp}(\omega) = \frac{1}{2}\left(d_x^{sp'}(\omega_+) + d_x^{sp'}(\omega_-) + id_y^{sp'}(\omega_+) - id_y^{sp'}(\omega_-)\right)$$
$$d_y^{sp}(\omega) = \frac{1}{2}\left(-id_x^{sp'}(\omega_+) + id_x^{sp'}(\omega_-) + d_y^{sp'}(\omega_+) + d_y^{sp'}(\omega_-)\right) \tag{A1}$$
$$d_z^{sp}(\omega) = d_z^{sp'}(\omega)$$

and the fluctuation-dissipation theorem (FDT) in the reference frame $\Sigma'(X',Y',Z')$ of rotating particle

$$\left\langle d_i^{sp'}(\omega') d_k^{sp'}(\omega) \right\rangle = 2\pi \delta_{ik} \delta(\omega+\omega')\hbar \alpha''(\omega)\coth\frac{\hbar\omega}{2k_BT}, \tag{A2}$$

we obtain

$$\left\langle d_x^{sp}(\omega) d_y^{sp}(\omega') \right\rangle = \frac{1}{4}\left\{ \begin{array}{l} \left(d_x^{sp'}(\omega_+) + d_x^{sp'}(\omega_-)\right)\left(-id_x^{sp'}(\omega'_+) + id_x^{sp'}(\omega'_-)\right) + \\ + \left(id_y^{sp'}(\omega_+) - id_x^{sp'}(\omega_-)\right)\left(d_y^{sp'}(\omega'_+) + d_x^{sp'}(\omega'_-)\right) \end{array} \right\} =$$
$$= \frac{i}{2}\left\{\left\langle d_x^{sp'}(\omega_+) d_x^{sp'}(\omega'_-) \right\rangle - \left\langle d_x^{sp'}(\omega_-) d_x^{sp'}(\omega'_+) \right\rangle \right\} = \tag{A3}$$
$$= \frac{i}{2} 2\pi\hbar\delta(\omega+\omega')\left[\alpha''(\omega_+)\coth\frac{\hbar\omega_+}{2k_BT} - \alpha''(\omega_-)\coth\frac{\hbar\omega_-}{2k_BT}\right]$$

$$\left\langle d_y^{sp}(\omega) d_x^{sp}(\omega') \right\rangle = -\left\langle d_x^{sp}(\omega) d_y^{sp}(\omega') \right\rangle =$$
$$= -\frac{i}{2} 2\pi\hbar\delta(\omega+\omega')\left[\alpha''(\omega_+)\coth\frac{\hbar\omega_+}{2k_BT} - \alpha''(\omega_-)\coth\frac{\hbar\omega_-}{2k_BT}\right] \tag{A4}$$

$$\left\langle d_x^{sp}(\omega) d_x^{sp}(\omega') \right\rangle = \left\langle d_y^{sp}(\omega) d_y^{sp}(\omega') \right\rangle =$$
$$= \frac{1}{2} 2\pi\hbar\delta(\omega+\omega')\left[\alpha''(\omega_+)\coth\frac{\hbar\omega_+}{2k_BT} + \alpha''(\omega_-)\coth\frac{\hbar\omega_-}{2k_BT}\right] \tag{A5}$$

$$\left\langle d_x^{sp}(\omega) d_z^{sp}(\omega') \right\rangle = \left\langle d_y^{sp}(\omega) d_z^{sp}(\omega') \right\rangle = 0 \tag{A6}$$

$$\left\langle d^{sp}_z(\omega)d^{sp}_z(\omega')\right\rangle = 2\pi\hbar\delta(\omega+\omega')\alpha''(\omega)\coth\frac{\hbar\omega}{2k_BT} \tag{A7}$$

It is worth noting that we omitted index "1" in the components of dipole moment.

b) $\boldsymbol{\Omega}=(\Omega,0,0)$ (Fig. 2)

In the same way, we obtain expressions for the correlators of the dipole moment differing from (A3)—(A7) by cyclic permutation $x\to y\to z\to x$

$$\left\langle d^{sp}_y(\omega)d^{sp}_z(\omega')\right\rangle = -\left\langle d^{sp}_z(\omega)d^{sp}_y(\omega')\right\rangle =$$
$$= \frac{i}{2}2\pi\hbar\delta(\omega+\omega')\left[\alpha''(\omega_+)\coth\frac{\hbar\omega_+}{2k_BT}-\alpha''(\omega_-)\coth\frac{\hbar\omega_-}{2k_BT}\right] \tag{A8}$$

$$\left\langle d^{sp}_y(\omega)d^{sp}_y(\omega')\right\rangle = \left\langle d^{sp}_z(\omega)d^{sp}_z(\omega')\right\rangle =$$
$$= \frac{1}{2}2\pi\hbar\delta(\omega+\omega')\left[\alpha''(\omega_+)\coth\frac{\hbar\omega_+}{2k_BT}+\alpha''(\omega_-)\coth\frac{\hbar\omega_-}{2k_BT}\right] \tag{A9}$$

$$\left\langle d^{sp}_y(\omega)d^{sp}_x(\omega')\right\rangle = \left\langle d^{sp}_z(\omega)d^{sp}_x(\omega')\right\rangle = 0 \tag{A10}$$

$$\left\langle d^{sp}_x(\omega)d^{sp}_x(\omega')\right\rangle = 2\pi\hbar\delta(\omega+\omega')\alpha''(\omega)\coth\frac{\hbar\omega}{2k_BT} \tag{A11}$$

**APPENDIX B**

In sections 3—5, we have calculated the contributions to the heating rate and frictional moment due to the vacuum fluctuations (irrespectively of the presence of the second particle) in line with the scheme used in [6]. In this case, the use is made of the limiting expression for the imaginary part of Green's function of photons in vacuum

$$\lim_{|\mathbf{r}-\mathbf{r}'|\to 0}\operatorname{Im}D_{ik}(\omega,\mathbf{r},\mathbf{r}') = \frac{2}{3}\frac{\omega^3}{c^3}\left(-\frac{\hbar c^2}{\omega^2}\right)\delta_{ik}$$

In this section, we calculate the values of $\dot{Q}$ and $M_z$ in $(\omega,\mathbf{k})$-representation for the Green function [14]. Though this method is somewhat more cumbersome as compared to the calculation based on the use of the frequency representation of Green's function, it has the advantage in solving problems of a certain type.

So, the rate of heat exchange in the system particle-background is determined from

$$\dot{Q}^{vac} = \left\langle \mathbf{d}_1^{sp}\mathbf{E}^{in}\right\rangle + \left\langle \mathbf{d}_1^{in}\mathbf{E}^{sp}\right\rangle$$

Writing $\mathbf{d}_1^{sp}(t)$ as the frequency Fourier integral and $\mathbf{E}^{in}(\mathbf{r}_1,t)$ as the Fourier integral over the frequency and wave-vector, we obtain ( at $x_1 = y_1 = 0$ )

$$\dot{Q}_1^{vac} = \left\langle \mathbf{d}_1^{sp}(t)\mathbf{E}^{in}(\mathbf{r}_1,t) \right\rangle = \int \frac{d\omega}{2\pi} \frac{d\omega'}{2\pi} \frac{d^3k}{(2\pi)^3} (-i\omega')\exp(-i(\omega+\omega')t)\left\langle \mathbf{d}_1^{sp}(\omega)\mathbf{E}^{in}(\omega,\mathbf{k}) \right\rangle \tag{B1}$$

The induced vacuum field $\mathbf{E}^{in}(\omega,\mathbf{k})$ is determined from the equation for the Hertz vectors [13]

$$\left(\Delta + \frac{\omega^2}{c^2}\varepsilon(\omega)\right)\Pi(\omega,\mathbf{k}) = -\frac{4\pi}{\varepsilon(\omega)}\mathbf{d}^{sp}(\omega) \tag{B2}$$

where $\Delta = -k^2$ and $\varepsilon(\omega) = 1 + i \cdot 0 \cdot sign(\omega)$. From (B2) we obtain

$$\Pi(\omega,\mathbf{k}) = \frac{4\pi \mathbf{d}^{sp}(\omega)}{k^2 - \omega^2/c^2 - i \cdot 0 \cdot sign\omega} = 4\pi \mathbf{d}^{sp}(\omega)\left[P\left(\frac{1}{k^2-\omega^2/c^2}\right) + i\pi\delta(k^2-\omega^2/c^2)\right] \tag{B3}$$

$$\mathbf{E}^{in}(\omega,\mathbf{k}) = graddiv\bar{\Pi}(\omega,\mathbf{k}) - \Delta\Pi(\omega,\mathbf{k}) = -\mathbf{k}(\mathbf{k}\bar{\Pi}(\omega,\mathbf{k})) + k^2\Pi(\omega,\mathbf{k}) \tag{B4}$$

Using (B1), (B3),(B4) and Eqs. (A5), (A7) for the rotating dipole yields

$$\dot{Q}_1^{vac} = -\frac{2\hbar}{3\pi c^3}\int_0^\infty d\omega\omega^4\left(\alpha_1''(\omega)\coth\frac{\hbar\omega}{2k_BT_1} + \alpha_1''(\omega_+)\coth\frac{\hbar\omega_+}{2k_BT_1} + \alpha_1''(\omega_-)\coth\frac{\hbar\omega_-}{2k_BT_1}\right) \tag{B5}$$

To calculate the contribution $\dot{Q}_2^{vac} = \left\langle \mathbf{d}_1^{in}(t)\mathbf{E}^{sp}(\mathbf{r}_1,t) \right\rangle$ in the starting expression for $\dot{Q}^{vac}$, one should bear in mind that in the case $\mathbf{\Omega} = (0,0,\Omega)$ the relation between the induced dipole moment of a particle and spontaneous vacuum field in the reference frame $\Sigma'$, with allowance for (3)—(5) takes the form ( $\mathbf{r}_1 = (0,0,R)$ )

$$d_x^{in}(t) = \int \frac{d\omega}{2\pi}\frac{d^3k}{(2\pi)^3}\frac{1}{2}\left[\begin{array}{l}\alpha_1(\omega_+)\left(E^{sp}_x(\omega,\mathbf{k}) + iE^{sp}_y(\omega,\mathbf{k})\right) + \\ + \alpha_1(\omega_-)\left(E^{sp}_x(\omega,\mathbf{k}) - iE^{sp}_y(\omega,\mathbf{k})\right)\end{array}\right]\exp(-i\omega t) \tag{B6}$$

$$d_y^{in}(t) = \int \frac{d\omega}{2\pi}\frac{d^3k}{(2\pi)^3}\frac{1}{2}\left[\begin{array}{l}-i\alpha_1(\omega_+)\left(E^{sp}_x(\omega,\mathbf{k}) + iE^{sp}_y(\omega,\mathbf{k})\right) + \\ + i\alpha_1(\omega_-)\left(E^{sp}_x(\omega,\mathbf{k}) - iE^{sp}_y(\omega,\mathbf{k})\right)\end{array}\right]\exp(-i\omega t) \tag{B7}$$

$$d_z^{in}(t) = \int \frac{d\omega}{2\pi} \frac{d^3k}{(2\pi)^3} \alpha_1(\omega) E_z^{sp}(\omega, \mathbf{k}) \exp(-i\omega t) \tag{B8}$$

Using (B6)—(B8) yields

$$\dot{Q}_2^{vac} = \left\langle \mathbf{d}_1^{in}(t) \mathbf{E}^{sp}(\mathbf{r}_1, t) \right\rangle = \int \frac{d\omega d^3k}{(2\pi)^4} \frac{d\omega' d^3k'}{(2\pi)^4} (-i\omega) \exp(-i(\omega+\omega')t) \cdot$$
$$\cdot \left\{ \begin{array}{l} \frac{1}{2}(\alpha_1(\omega_+) + \alpha_1(\omega_-))\left[\left\langle E^{sp}_x(\omega,\mathbf{k})E^{sp}_x(\omega',\mathbf{k}')\right\rangle + \left\langle E^{sp}_y(\omega,\mathbf{k})E^{sp}_y(\omega',\mathbf{k}')\right\rangle\right] + \\ + \alpha_1(\omega)\left\langle E^{sp}_z(\omega,\mathbf{k})E^{sp}_z(\omega',\mathbf{k}')\right\rangle + (...) \end{array} \right\} \tag{B9}$$

The terms (…) in Eq. (B9) are omitted since they result in zero values after integration. Correlators in (B9) are evaluated using the FDT [2]

$$\left\langle E^{sp}_i(\omega,\mathbf{k})E^{sp}_k(\omega',\mathbf{k}')\right\rangle = (2\pi)^4 \delta(\omega+\omega')\delta(\mathbf{k}+\mathbf{k}') \frac{2\pi^2 \hbar}{k}\left(\frac{\omega^2}{c^2}\delta_{ik} - k_i k_k\right) \cdot$$
$$\cdot \left[\delta\left(\frac{\omega}{c} - k\right) - \delta\left(\frac{\omega}{c} + k\right)\right] \coth\frac{\hbar\omega}{2k_B T_3} \tag{B10}$$

Substituting (B10) into (B9) yields

$$\dot{Q}_2^{vac} = \frac{2\hbar}{3\pi c^3} \int_0^\infty d\omega \omega^4 \left(\alpha_1''(\omega_+) + \alpha_1''(\omega_-) + \alpha_1''(\omega)\right) \coth\frac{\hbar\omega}{2k_B T_3} \tag{B11}$$

Summing (B5) and (B11) we finally obtain

$$\dot{Q}^{vac} = -\frac{2\hbar}{3\pi c^3} \int_0^\infty d\omega \omega^4 \cdot$$
$$\cdot \left\{ \begin{array}{l} \alpha_1''(\omega)\left(\coth\frac{\hbar\omega}{2k_B T_1} - \coth\frac{\hbar\omega}{2k_B T_3}\right) + \\ + \alpha_1''(\omega_+)\left(\coth\frac{\hbar\omega_+}{2k_B T_1} - \coth\frac{\hbar\omega}{2k_B T_3}\right) + \alpha_1''(\omega_-)\left(\coth\frac{\hbar\omega_-}{2k_B T_1} - \coth\frac{\hbar\omega}{2k_B T_3}\right) \end{array} \right\} \tag{B12}$$

The frictional torque is calculated in the same way:

$$M_z = -\frac{2\hbar}{3\pi c^3} \int_0^\infty d\omega\, \omega^3 \begin{bmatrix} \alpha_1''(\omega_-)\left(\coth\frac{\hbar\omega_-}{2k_B T_1} - \coth\frac{\hbar\omega}{2k_B T_3}\right) - \\ \alpha_1''(\omega_+)\left(\coth\frac{\hbar\omega_+}{2k_B T_1} - \coth\frac{\hbar\omega}{2k_B T_3}\right) \end{bmatrix} \quad \text{(B13)}$$

When considerinng the second configuration $\mathbf{\Omega}=(\Omega,0,0)$, we obviously obtain the same expressions for $\dot{Q}$ and $M_x$.

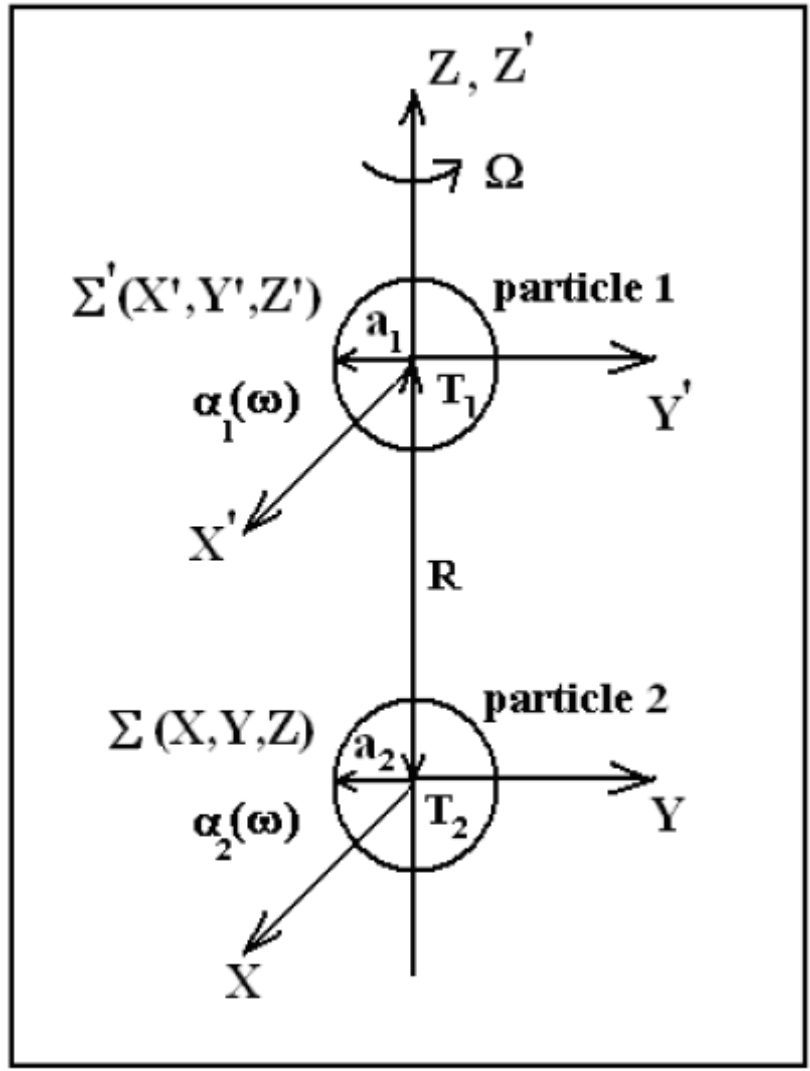

Fig. 1 Geometrical configuration 1 and coordinate systems used

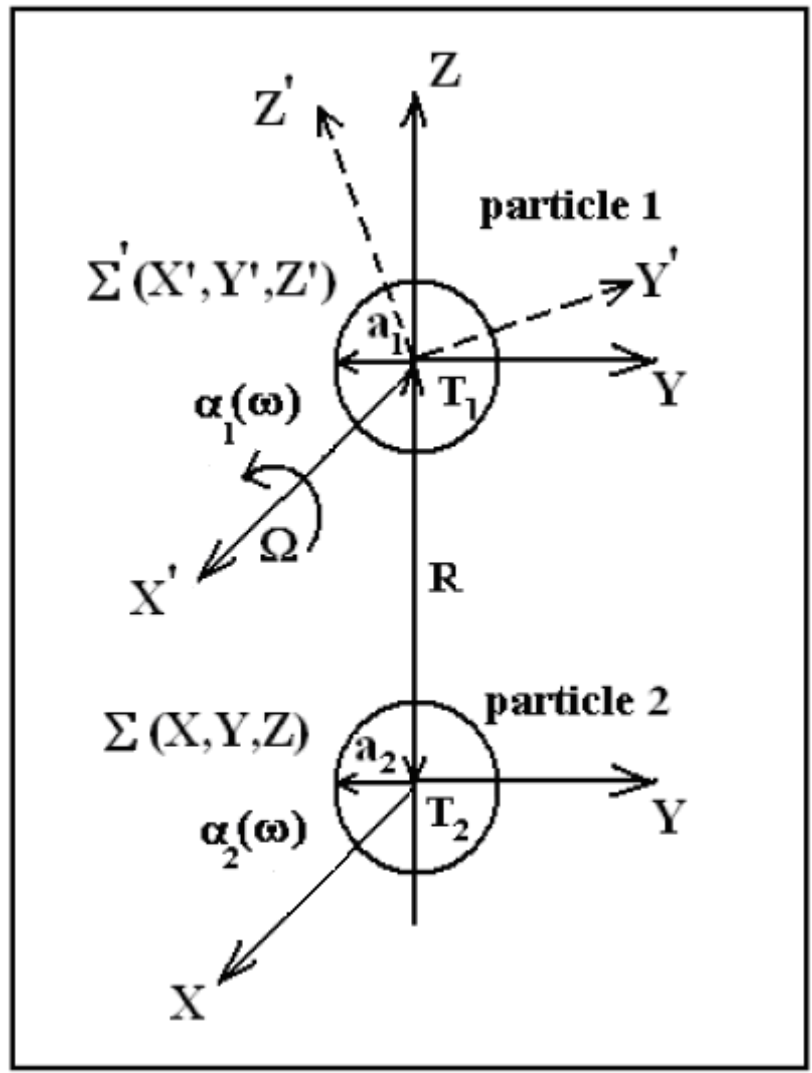

Fig. 2 Geometrical configuration 2 and coordinate systems used